\documentclass[aps,prl,twocolumn,showpacs,amsmath,amssymb,superscriptaddress,tightenlines]{revtex4}
\usepackage{graphicx}

\begin{document}
\date{\today}

\title{Vorticity and Magnetic Field Generation from Initial Anisotropy in Ultrarelativistic Gamma-Ray Burst Blastwaves}

\author{Milo\v s Milosavljevi\'c}
\affiliation{University of Texas, 1 University Station C1400, Austin, TX 78712}
\author{Ehud Nakar}
\affiliation{California Institute of Technology 130-33, 1200 East California Boulevard, Pasadena, CA 91125}
\author{Fan Zhang}
\affiliation{University of Texas, 1 University Station C1400, Austin, TX 78712}

\begin{abstract}

Because conical segments of quasispherical ultrarelativistic blastwaves are causally disconnected on angular scales larger than the blastwave inverse Lorentz factor, astrophysical blastwaves can sustain initial anisotropy, imprinted by the process that drives the explosion, while they remain relativistic.  We show that initial angular energy fluctuations in ultrarelativistic blastwaves imply a production of vorticity in the blastwave, and calculate the vortical energy production rate. In gamma-ray burst (GRB) afterglows, the number of vortical eddy turnovers as the shocked fluid crosses the blastwave shell is about unity for marginally nonlinear anisotropy. Thus the anisotropy must be nonlinear to explain the magnetic energy density inferred in measured GRB spectra.

\end{abstract}

\pacs{91.30.Mv, 98.70.Rz}

\maketitle

Special relativistic shocks occur where the gravity of a compact object,
such as a black hole, accelerates plasma to ultrarelativistic
velocities. Emission form such shocks is observed in astrophysical
sources on many length, time, and energy scales, and is often attributed
to the synchrotron process. The composition of the plasma, the magnitude
of the preshock magnetic field, and the Lorentz factor of the shock are
generally unknown.  In the best-studied system---the radiative afterglow
that follows the bright flash of $\gamma$-rays in gamma-ray burst (GRB)
phenomenon---the structure of the emitting region can be modeled from
the observed emission. In the standard GRB afterglow model
\cite{Piran:05a,Nakar:07}, the radiation is produced in a relativistic
blastwave shell propagating into a weakly magnetized plasma. The
afterglow emission is then the synchrotron radiation from nonthermal
electrons gyrating in a strong magnetic field of the shocked plasma.
Detailed studies of GRB spectra and light curves have shown that the
magnetic energy density in the emitting region is a fraction
$\epsilon_B\sim 10^{-2}$ to $10^{-3}$ of the internal energy density
\cite{Panaitescu:02} and that the magnetic field must be present over at
least a few percent of the blastwave thickness \cite{Rossi:03}. The
origin of this downstream magnetic field is a longstanding important
open question.
 
Compressional amplification of the weak pre-existing microgauss magnetic field of
the circumburst medium merely yields
$\epsilon_B\sim 10^{-9}$ \cite{Gruzinov:01}. The strong small-scale field generated in
collisionless plasma instabilities in the shock transition
was considered as a candidate for the post-shock
field \cite{GruzinovWaxman:99}. However, simulations of  shock
transition layers \cite{Spitkovsky:07} indicate that the small-scale field
decays rapidly over few plasma skin depths ($\lambda_{\rm s}$), and does
not persist on distances from the shock transition
($\sim10^9\lambda_{\rm s}$) where the emission originates. Suggestions
have been made that a persistent magnetic field develops in the shock
precursor due to the streaming of shock-accelerated protons
\cite{Milosavljevic:06} or cascade-generated $e^\pm$ pairs
\cite{RamirezRuiz:07}. The feasibility of these scenarios depends
on the insufficiently understood details of particle acceleration and
streaming  instabilities.
 
Recently, Sironi \& Goodman \cite{Goodman:07} showed that if the
unshocked medium is strongly inhomogeneous, significant vorticity is
produced in the shock transition. The resulting downstream turbulence
amplifies the weak seed magnetic field of the preshock medium to the
observed level. The density of the interstellar medium may vary at the
required level in GRB associated with the iron core collapse in
mass-losing very massive stars (i.e., in ``long''-type GRBs), but is not
expected in the GRB associated with old progenitors (i.e., in
``short''-type GRBs). While the strength of the magnetic field in the
afterglows of short-type GRBs is not well-constrained, the similarity of
the afterglow light curves and spectra in the long- and short-type GRBs
\cite{Nakar:07} suggests that the magnetic field is amplified by the
same process in both systems.

Here we propose a source of vortical energy in the shock downstream that
could generically be present in any ultrarelativistic blastwave: the
initial angular anisotropy of the energy carried by the blastwave. A
unique feature of ultrarelativistic blastwaves is that they are composed
of many small, causally-disconnected patches. The patches lose causal
contact when the driving outflow (e.g., an electromagnetic jet launched
near a black hole) accelerates the swept-up ambient medium, and evolve
independently until the blastwave decelerates sufficiently (see below). 
In GRBs, an additional source of angular inhomogeneities is the phase in
which the initial (``prompt'') $\gamma$-rays are emitted, when a sizable
fraction of the outflow energy is dissipated and emitted as
$\gamma$-rays.  This takes place after the relativistic outflow achieves
its maximal Lorentz factor, and thus any variation in the dissipation in
causally disconnected regions translates directly into angular energy
fluctuations in the shock wave that ultimately blasts into the ambient
medium.
 
There is also direct observational motivation for 
initial angular anisotropy in GRB outflows.
Significant angular fluctuations have been invoked to
explain the large variation of the $\gamma$-ray luminosity between
bursts \cite{Kumar:00} and the intraburst variability observed in many
afterglows \cite{Nakar:03}.   
The afterglow polarization \cite{Covino:04}
indicates a breaking of axial symmetry; the correlatedness of this
polarization with large amplitude variability suggests blastwave anisotropy 
as the source of variability \cite{Granot:03,Nakar:04}.
E.g., the afterglow in GRB 021004
exhibits a strong light curve variability correlated with a variable
linear polarization; anisotropy on scales $\sim 2\pi/l$ with $l
\sim 200$ and a nonlinear energy contrast $\sim 3$ can
explain the data \cite{Nakar:04}. 

We proceed to estimate the conditions for successful turbulent magnetic field
amplification in
quasispherical ultrarelativistic blastwaves with initial angular
energy fluctuations. Consider an ultrarelativistic point explosion with
total isotropic-equivalent energy $E$ propagating into a medium of
uniform density $\rho_0$.  We work in the rest frame of the unshocked
fluid and the explosion center.  In spherical symmetry, the Lorentz
factor of the strong shock wave that forms at the leading edge of the
blastwave decays in time $t$ as $\Gamma=(\frac{17}{8\pi}E
/t^3\rho_0)^{1/2}$ and its position is located at
$R=t(1-\frac{1}{8}\Gamma^{-2})$ (the speed of light is unity). 
The Lorentz factor in the shock downstream $\gamma=\Gamma/(2\chi)^{1/2}$
and the pressure $p=\frac{2}{3}\Gamma^2/\chi^{17/12}$ are expressed in
terms of the Blandford-McKee \cite{Blandford:76} 
variable $\chi(r,t)= (t-r)/(t-R)+{\cal O}(\Gamma^{-2})$.  Here, $r$ is
the radius from the center of the explosion and $\Gamma\gg 1$ is assumed.

Consider initial ($t\rightarrow 0$) linear fluctuations in the energy
of the explosion, $E(\theta,\phi)=[1+\delta_E(\theta)]E$, where at no
loss of generality we assume axial symmetry, $\partial_\phi\rightarrow
0$ and $\delta_E(\theta)=\delta_E Y_{l 0} (\theta)$. Here
$Y_{lm}(\theta,\phi)$ is the spherical harmonic; the case with an
azimuthal dependence, $m\neq0$, can be treated equivalently. Conical
angular segments of the blastwave separated by
$\Delta\theta\gtrsim\frac{2}{3}\Gamma^{-1}$, which corresponds to
$l\lesssim3\pi\Gamma$, have not been in causal contact since the
beginning of the explosion, and are evolving independently as fragments
of spherical explosions with energy fluences displaced from the
spherical average.  Following \cite{Gruzinov:00}, we expand the radial
Lorentz factor $\gamma(r,\theta,t)=\gamma(r,t)[1+\delta_\gamma(r,t)
Y_{l0}(\theta)]$, the pressure $p(r,\theta,t)=p(r,t)[1+\delta_p(r,t)
Y_{l0}(\theta)]$, the shock radius $R(\theta,t)=R(t)[1+\delta_R(t)
Y_{l0}(\theta)]$, and the velocity of the fluid in the $\hat\theta$
direction $u(r,\theta,t)=u(r,t) \partial_\theta Y_{l0}(\theta)$, which
is here assumed to be Newtonian, as linear perturbations around their
spherical averages.  Then, at times $t\gg t_{\rm crit}\equiv (\frac{153
\pi}{8} E/ l^2\rho_0)^{1/3}$, i.e., when
$\Delta\theta>\frac{2}{3}\Gamma^{-1}$ and for $\chi(r,t)\sim 1$, the
fluctuations simply inherit their initial values and are independent
of time and radius, $\delta_p=-\frac{5}{12}\delta_E$,
$\delta_R=\frac{1}{8}\Gamma^{-2}\delta_E$, $\delta_\gamma=0$, and $u=0$. 

Gruzinov \cite{Gruzinov:00} linearizes the equations of relativistic
hydrodynamics, $\partial_\alpha T^{\alpha\beta}=0$, where
$T^{\alpha\beta}$ is the energy-momentum tensor, and changes variables
$(r,t)\rightarrow (\xi,\tau)$, where $\xi\equiv\frac{1}{4}\ln\chi(r,t)$
and $\tau\equiv -\frac{2}{3}\ln\Gamma(t)$, to derive equations governing
the evolution of the fluctuations
\begin{eqnarray}
\label{eq:fluid}
\dot\delta_\gamma +3\delta_\gamma'-\frac{3}{2}\delta_p'-\delta_\gamma+\frac{1}{2}l(l+1) u&=&0 ,\nonumber
\end{eqnarray}
\begin{eqnarray}
\dot \delta_p + 3\delta_p' - 8\delta_\gamma'+\frac{40}{3}\delta_\gamma-2l(l+1)u&=&0 ,\nonumber\\
\dot u+u'-\frac{14}{3} u+\frac{1}{2} e^{3\tau+4\xi}\delta_p &=&0 ,
\end{eqnarray}
where the overdot and the prime denote derivatives with respect to
$\tau$ and $\xi$, respectively, and the equations are valid for $\xi\ll
-\frac{3}{4}\tau$ and $\tau<0$, as the blastwave is Newtonian at larger
$\xi$ or $\tau$.  At the shock transition ($\xi=0$), the continuity of
the energy-momentum flow across the transition implies the shock jump
conditions \cite{Gruzinov:00}
\begin{equation}
\label{eq:boundary}
\dot\delta_p=2\dot\delta_\gamma-\frac{10}{3}\delta_\gamma, \ \ \ \  
u=\frac{3}{10} e^{3\tau}(\delta_p - 2\delta_\gamma) .
\end{equation}
The first two of Eq.~\ref{eq:fluid} can be expressed in terms of the Riemann variables $f_\pm\equiv 2\delta_\gamma\pm\frac{\sqrt{3}}{2}\delta_p$ propagating along the $C^\pm$ characteristics with velocities $c^\pm= 3\mp 2\sqrt{3}$ \cite{Gruzinov:00}
\begin{eqnarray}
\label{eq:riemann}
\dot f^\pm+c^\pm (f^\pm)' - \alpha^\pm(f^++f^-)+\zeta^\pm u&=&0 ,
\end{eqnarray} 
where $\alpha^\pm\equiv \frac{1}{2}\mp\frac{5}{\sqrt{3}}$ and $\zeta^\pm\equiv (1\mp\sqrt{3})l(l+1)$.
Following \cite{Gruzinov:00}, we adopt the notation $f\equiv f_+$ and $g\equiv f_-$.

Dynamics of the shocked fluid governed by Eq.~\ref{eq:fluid} is easily
understood: an initial energy-pressure fluctuation $\delta_p$ sources
transverse ($\delta_p,u$) and longitudinal ($\delta_p,\delta_\gamma$)
acoustic fluctuations, which oscillate after the mode comes within the
causal horizon at $t=t_{\rm crit}$ ($\delta_\gamma$ equals the radial
velocity perturbation in the local fluid rest frame).  This is analogous
to the horizon re-entry of superhorizon cosmological fluctuations in the
early universe. 

The initial value problem (Eq.~\ref{eq:fluid}) subject to the
boundary conditions (Eq.~\ref{eq:boundary}) is well-posed if
a boundary condition at the origin of the $C^+$ characteristics, at some
distance $\xi_{\rm max}$ from the shock is specified. Gruzinov
\cite{Gruzinov:00} reports that the solution near the
shock is independent of the boundary condition if $\xi_{\rm max}$ is
large enough.  The independence
can be understood as follows.  Since $\alpha^+/c^+>0$, the fluctuations
$f^+$ are damped from large to small $\xi$; this
is because the explosion energy is concentrated near the
shock \cite{Blandford:76}.  The reverse $f^-$ fluctuations grow 
much slower than the $f^+$ decay, so the former do not seed the latter
at large $\xi$. Having specified a
boundary condition, e.g., $f(\xi_{\rm max})=0$, 
the evolution of perturbations in the shock
downstream is uniquely determined and can be solved for, e.g., after
Taylor expansion $\psi(\xi,\tau)=\sum_{n=0}^\infty (n!)^{-1}
\xi^n \psi_n(\tau)$, where $\psi=f,g,u$. 

An approximate solution is obtained after assuming that the
perturbations decay after they start to oscillate, and that the decay is
complete everywhere except within few wavelengths
$N\lambda^+$ from the shock, where $\lambda^+=2\pi |c^+|/\kappa^+$, and
$\kappa^+$ is the frequency of oscillations of fluctuations
propagating along the $C^+$ characteristic.  The frequency is estimated
by substituting the last of Eq.~\ref{eq:fluid} in the first of
Eq.~\ref{eq:boundary} which yields $\kappa^+\approx
[\frac{1}{2}(1-\frac{1}{\sqrt{3}})l(l+1)]^{1/2}e^{3\tau/2}$.  Thus we
set $\xi_{\rm max}\rightarrow N \lambda^+ \propto e^{-3\tau/2}$; the
decay proceeds from large to small distances from the shock. Expanding
to the linear order yields
\begin{eqnarray}
\label{eq:linear}
\dot f_0 + c^+ f_1 - \alpha^+(f_0+g_0)+\zeta^+u_0&=&0 ,\nonumber\\
\dot g_0 + c^- g_1 - \alpha^-(f_0+g_0)+\zeta^-u_0&=&0 ,\nonumber\\
\dot u_0 + u_1 -\frac{14}{3} u_0 + \frac{1}{2} e^{3\tau} \left(\frac{f_0-g_0}{\sqrt{3}}\right)&=& 0 ,\nonumber\\
\frac{\dot f_0-\dot g_0}{\sqrt{3}} - \frac{\dot f_0 + \dot g_0}{2}
+ \frac{5}{3}  \frac{f_0 + g_0}{2}  &=& 0 ,\nonumber\\
f_0 + \xi_{\rm max}(t) f_1 &=& 0 , \nonumber\\
u_0 - \frac{3}{10} e^{3\tau} \left(\frac{f_0-g_0}{\sqrt{3}}-\frac{f_0+g_0}{2}\right) &=& 0 ,
\end{eqnarray}
where the first three equations are the original equations of motions
for $f$, $g$, and $u$, and the last three are the two constraints
imposed at $\xi=0$ (the shock jump conditions), and a constraint imposed
at $\xi=\xi_{\rm max}$. Integrating the ODEs in Eq.~\ref{eq:linear} with
initial conditions $f_0(-\infty)=-g_0(-\infty)$ and $u_0(-\infty)=0$ and $N=1$, we
find that the resulting approximate solution
$\delta_p(\xi=0,\tau)=\frac{1}{\sqrt{3}}[f_0(\tau)-g_0(\tau)]$ closely
matches the solution in \cite{Gruzinov:00}, thus
validating our approximations.

We employ the definition of relativistic vorticity $\vec{\omega}\equiv
\vec{\nabla}\times\vec{H}$ \cite{Goodman:07}, which is conserved,
$\partial_t\vec{\omega}-\vec{\nabla}\times(\vec{v}\times\vec{\omega})=0$, in
the smooth (i.e., shock-free) part of the flow in an ideal, barotropic
fluid with ultrarelativistic equation of state. Here, $\vec{H}\equiv
p^{1/4}\gamma \vec{v}$, where as before, $p$ is the fluid pressure,
$\gamma=(1-v^2-u^2)^{-1/2}$ is its Lorentz factor, and we have
decomposed fluid velocity into $\vec{v}=v\hat{r}+u\hat{\theta}$. Since
$u\ll v$, $\gamma\approx(1-v^2)^{-1/2}$ is the radial Lorentz factor. 
Vorticity vanishes in an unperturbed, spherically-symmetric blastwave,
but in a blastwave with initial anisotropy, it is produced at the
shock and advected into the downstream.  In the
$(\xi,\tau)$ coordinates, the vorticity conservation equation becomes
$\dot\omega+\omega'+9\omega+{\cal O}(\Gamma^{-2})=0$, where $\omega$ is
the amplitude of the sole nonvanishing component of the vorticity,
$\vec{\omega}=\omega \partial_\theta Y_{l0}\hat\phi$.  The vorticity
decreases into the downstream,
$\omega(\xi,\tau)=e^{-9\xi}\omega(0,\tau-\xi)$.  In terms of the fluid
perturbation the vorticity equals
\begin{equation}
\label{eq:omega_perturb}
\omega\approx\frac{p^{1/4}\gamma}{t} \left[\left(\frac{41}{6}u
-2u'\right)e^{-3\tau-4\xi}-\delta_\gamma-\frac{1}{4}\delta_p\right] ,
\end{equation}
where we have kept only the leading-order terms. The vorticity in
Eq.~\ref{eq:omega_perturb} is conserved if the fluid perturbations evolve
according to Eq.~\ref{eq:fluid}.  Because
$\delta_p(\tau\rightarrow-\infty)\neq 0$ while $\delta_\gamma\sim 0$ and
$u\sim0$, vorticity does not vanish identically prior to the horizon
entry of the mode.

We follow the prescription in \cite{Goodman:07} to calculate the
vortical energy, which is defined in the local, noninertial rest frame
of the shocked fluid; we denote the quantities in this frame with tilde,
${\tilde r}=\gamma(r-v t)$ and ${\tilde t}=\gamma(t-v r)$. Vorticity
transforms as a $2$-form $\Omega\equiv \omega_\phi dr\wedge d(r\theta)$.
Since $dr\wedge d(r\theta)\rightarrow \gamma^{-1} d\tilde r\wedge
d(r\theta)$ to the leading order in $\gamma^{-1}$ (terms involving
$d\tilde t$ do not contribute to vorticity), we have
$\tilde\omega\approx\omega/\gamma$, and the $\xi$-dependence of the
vorticity is given by $\tilde
\omega(\xi,\tau)=e^{-(11/2)\xi}\tilde\omega(0,\tau-\xi)$. 

The fluid is subject to differential acceleration, and thus there is not
a single inertial frame in which we can calculate the vortical energy. 
As an approximation, we work in the instantaneous inertial rest frame of
the fluid element at $\xi=0$.  Simultaneity in this frame is equivalent
to $d\tilde t(\xi,\tau)=0$, which implies that $d\tilde r\approx
\frac{\sqrt{2}}{3} \Gamma^{-1} t d\xi$.  This is not exact; the
numerical coefficient depends on the approximation, 
and thus the forthcoming estimates are crude.
We also replace covariant derivatives with normal derivatives in what
follows. We carry out approximate projection of the vorticity onto the
shock plane
\begin{equation}
\tilde \sigma = \int \tilde\omega d\tilde r \approx \frac{t}{3\gamma} \int_0^\infty\tilde\omega d\xi \approx  \frac{2}{33}\frac{t}{\gamma} \tilde\omega(\xi=0) .
\end{equation}

\begin{figure}
\includegraphics[width=3.0in]{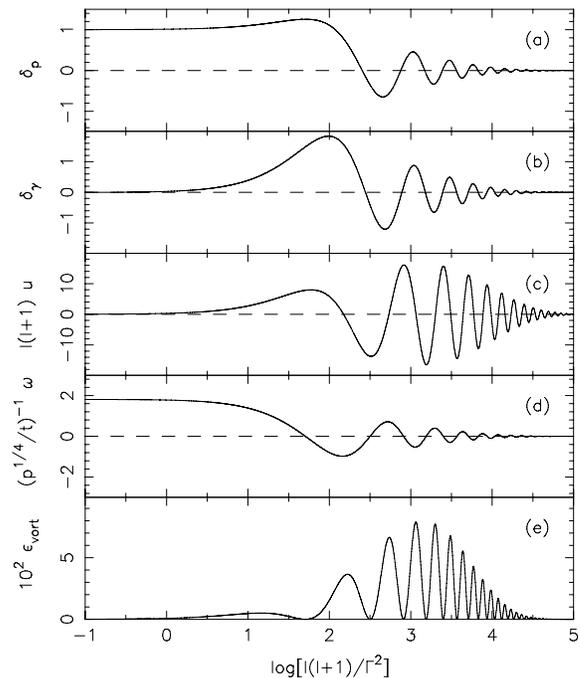}
\caption{Evolution of the pressure perturbation $\delta_p$, radial
Lorentz factor perturbation $\delta_\gamma$, transverse velocity $u$,
and the fluid-frame vorticity $\tilde\omega$, and the fractional energy
in vortical motions at the shock.  The quantities $\delta_p$,
$\delta_\gamma$, and $\tilde\omega$ are expressed in the units of the
initial pressure anisotropy $\delta_{p,0}$, and $\epsilon_{\rm vort}$ is
expressed in the units of $\delta_{p,0}^2$.}
\label{fig:figure}
\end{figure}

The vortical energy density in the local fluid rest frame equals
\cite{Goodman:07} $\tilde\varepsilon_{\rm vort}=\frac{4}{3} \rho
|\vec{\tilde H}_{\rm sol}|^2p^{-1/2}$, where $\vec{\tilde H}_{\rm sol}$
is the solenoidal component of $\vec{\tilde H}$, and $\rho$ is the
proper energy density. The solenoidal component can be written as a curl
of a vector potential $(\vec{\tilde H})_{\rm sol}=\vec{\tilde
\nabla}\times\vec{A}$.  Setting
$\vec{\tilde\nabla}\cdot\vec{A}=0$, the vorticity is related to
the potential via $\tilde
\nabla^2\vec{A}=-\vec{\tilde\omega}$.  Recall that $\vec{\tilde
\omega}\propto \partial_\theta Y_{l0}$, which is not an eigenfunction of
the angular component of the Laplacian.  However, for $l\gg 1$, at no
loss of generality we can restrict analysis to a small patch of the
spherical shell, e.g., near the equator, $|\theta-\frac{\pi}{2}|\lesssim
l^{-1}$, where $Y_{l0}(\theta)\approx \frac{1}{\pi} \cos(l\theta)$ and
the shell is locally quasi-planar.  Then $\vec{A}=
A\sin(l\theta)\hat\phi$ and $A=\frac{1}{2\pi} t\tilde\sigma e^{-l\tilde
r/t}$ is the solution of (we freely set
the origin of the coordinate $\tilde r=0$ at $\xi=0$)
\begin{equation}
\frac{\partial^2 A}{\partial\tilde r^2}-\frac{l^2}{t^2}
A=\frac{l}{\pi}  \tilde \omega(\tilde r,\tilde t) \approx \frac{l}{\pi}\tilde\sigma \delta(\tilde r) ,
\end{equation}
where $\delta(\tilde r)$ is the $\delta$-function.  
Close to the shock we take curl to obtain
$\langle ({\tilde H}_{\rm sol})^2\rangle\approx (\frac{1}{2\pi}l\tilde \sigma)^2
\approx (\frac{1}{33\pi}l t\tilde\omega/\gamma)^2$,
where the angular brackets denote an average over $\cos\theta$.  Substituting in 
$\varepsilon_{\rm vort}$ and diving by the internal energy $\rho$ yields the fractional
vortical energy $\epsilon_{\rm vort}\equiv \varepsilon_{\rm vort}/\rho$
\begin{eqnarray}
\epsilon_{\rm vort}&\approx& K \frac{l^2}{\Gamma^2} 
\left[\left(\frac{41}{6}u
-2u'\right)e^{-3\tau-4\xi}-\delta_\gamma-\frac{1}{4}\delta_p\right]^2 .
\end{eqnarray}
where $K= \frac{8}{3}\left(\frac{1}{33\pi}\right)^2\approx2.5\times10^{-4}$ is a numerical constant sensitive to the specifics of our approximations.

Fig.~\ref{fig:figure}, shows the evolution of the fluid
perturbation variables, the fluid-frame vorticity and the fractional
energy in vortical motions.  The latter oscillates
and reaches peak average amplitude $\epsilon_{\rm vort}\sim
3\times10^{-2} {\delta_{p,0}}^2$ after an $l$-mode becomes causal, for
$\Gamma\sim (0.01-0.1)\times l$.  Here,
$\delta_{p,0}\equiv\delta_p(\Gamma=\infty)$ is the initial fractional
pressure fluctuation.  The plots in Fig.~\ref{fig:figure} are scaled to
be independent of $l$, but abscissal range satisfying the premise that
the blastwave is ultrarelativistic does depend on $l$.

A fluid with nonvanishing vorticity develops eddies
which give rise to magnetic field amplification via the turbulent dynamo
mechanism \cite{Meneguzzi:81}.  Strong amplification is expected when
the number of eddy turnovers is larger than unity.
The number is estimated as the ratio $N_{\rm eddy}\sim t_{\rm
cross}/t_{\rm eddy}$, of the time scale $t_{\rm cross}\sim \frac{1}{4}t/\gamma$ 
on which the
fluid crosses the shell of the blastwave to the eddy turnover time
$t_{\rm eddy }\sim p^{1/4}\langle |\vec{\tilde\omega}|^2\rangle^{-1/2}$.  Using
$\langle(\partial_\theta Y_{l0})^2\rangle=\frac{1}{4\pi}l(l+1)$, we have
\begin{equation}
N_{\rm eddy}\sim \frac{1}{4\sqrt{2\pi}} \frac{l}{\Gamma} \frac{t\tilde\omega}{p^{1/4}} \approx 6.3\ \sqrt{\epsilon_{\rm vort}} \sqrt{\frac{2\times10^{-4}}{K}} .
\end{equation}
For $\epsilon_{\rm vort}\sim0.03$, we expect $N_{\rm eddy}\sim 1$.  
Therefore, within the limits of our 
approximations,
a nonlinear initial
perturbation, $\delta_{p,0}\gtrsim 1$, is necessary to amplify the magnetic field.
If amplification happens, the final energy density in the
field will be in approximate equipartition with the vortical
motions, $\epsilon_B\sim\epsilon_{\rm vort}$.  
With $\delta_{p,0}\sim1$ we find that $\epsilon_B\sim 3\times
10^{-2}$, consistent with the typical values inferred from GRB
spectra.  The anisotropy of GRB
blastwaves could be fully nonlinear. Since our
treatment is inadequate in the nonlinear regime 
(see \cite{Gruzinov:07} for an analytic approach to aspherical blastwaves),
numerical simulations of relativistic blastwave hydrodynamics are needed
to study the nonlinear turbulence produced in the shock.

Fig.~\ref{fig:figure} shows that the vorticity associated with an
$l$-mode persists through an increase by $\times10^3$ in $\Gamma^2$,
which corresponds to an increase by $\times10^4$ in the observer time
($t_{\rm obs} \propto \Gamma^{-8/3}$). Thus a single mode $l\sim 500$
produces vorticity during the observed GRB afterglow ($10^2-10^6\textrm{
s}$ after the burst). Since $\Gamma_{\rm max} \sim 100$, we indeed
expect such an $l$ mode from causality considerations.  Our estimates
indicate that $\delta_E\sim 1$ is required for strong magnetic field
amplification, although in this regime, our linear treatment is not
strictly applicable.  A measurable signature of such angular anisotropy
is afterglow flux variability at the level $\sim \delta_E$ on time
scales $t_{\rm obs}$ for which $l \sim 2\pi\Gamma$ \cite{Nakar:03}. At
later times the variability amplitude decays as $\sim
2\pi(\Gamma/l)\delta_E$ and can drop below $1\%$ while the vortical
energy is still high ($l/\Gamma \sim 100$). The observed variability
depends on the initial spectrum of the fluctuations. In many GRB
afterglows a variability is observed as predicted
in our model. In some cases no flux variability is observed around
$10^5\textrm{ s}$ after the burst down to the level $1\%-10\%$. There, as we
explain above, large-$l$ fluctuations may be responsible for magnetic
field generation without imprinting an observable variability.


\begin{thebibliography}{}

\bibitem[Piran(2005)]{Piran:05a} 
T.\ Piran,\ Rev.\ Mod.\ Phys., {\bf 76}, 114 (2005);
P.\ M\'esz\'aros, Rep. Prog. Phys., {\bf 69}, 2259 (2006).

\bibitem[Nakar(2007)]{Nakar:07} E.\ Nakar, Phys.\ Rep., {\bf 442},
166 (2007).

\bibitem[Panaitescu and Kumar(2002)]{Panaitescu:02} A.\ Panaitescu, and P.\ Kumar, Astrophys. J., {\bf 571}, 779 (2002);
S.~A.\ Yost, F.~A.\ Harrison, R.\ Sari, and D.~A.\ Frail, Astrophys. J., {\bf 597}, 459 (2003).

\bibitem[Rossi and Rees(2003)]{Rossi:03} E.\ Rossi, and M.~J.\ Rees, 
Mon. Not. R. Astr. Soc., {\bf 339}, 881 (2003).


\bibitem[Gruzinov(2001)]{Gruzinov:01} A.\ Gruzinov, Astrophys. J., {\bf 563}, L15 (2001).

\bibitem[Gruzinov and Waxman(1999)]{GruzinovWaxman:99} 
A.\ Gruzinov, and E.\ Waxman, Astrophys. J., {\bf 511}, 852 (1999);
M.~V.\ Medvedev, and A.\ Loeb, Astrophys. J., {\bf 526}, 697 (1999);
J.~T.\ Frederiksen, C.~B.\ Hededal, T.\ Haugb{\o}lle, and 
{\AA}.\ Nordlund, Astrophys. J., {\bf 608}, L13 (2004); 

\bibitem[Spitkovsky(2007)]{Spitkovsky:07} 
A. Spitkovsky, arXiv:0706.3126 (2007).


\bibitem[Milosavljevi{\'c} and Nakar(2006)]{Milosavljevic:06} 
M.\ Milosavljevi{\'c}, and E.\ Nakar, Astrophys. J., {\bf 651}, 979 (2006).

\bibitem[Ramirez-Ruiz et al.(2007)]{RamirezRuiz:07} E.\ Ramirez-Ruiz, 
K.-I.\ Nishikawa, and C.~B.\ Hededal, arXiv:0707.4381 (2007).





\bibitem[Kumar and Piran(2000)]{Kumar:00} P.\ Kumar, and T.\ Piran, Astrophys. J., {\bf 535}, 152 (2000).
 
\bibitem[{Nakar}, {Piran} and {Granot} (2003)]{Nakar:03}
E.\ Nakar, T.\ Piran, and J.\ Granot, New Astron., {\bf 8}, 495 (2003);
K.\ Ioka, S.\ Kobayashi, and B.\ Zhang, Astrophys. J., {\bf 631}, 429 (2005).

\bibitem[Covino et al.(2004)]{Covino:04} S.\ Covino, G.\ Ghisellini, D.\ Lazzati, and D.\ Malesani, Astronom. Soc.
Pacific Conf. Ser., {\bf 312}, 169 (2004).

\bibitem[Granot and K\"onigl(2003)]{Granot:03} J.\ Granot, and A.\ K\"onigl, Astrophys. J., {\bf 594}, L83 (2003).
 
\bibitem[{Nakar} and {Oren}(2004)]{Nakar:04}
E.\ Nakar, and Y.\ Oren, Astrophys. J., {\bf 602}, L97 (2004).


\bibitem[Blandford and McKee(1976)]{Blandford:76} R.~D.\ Blandford,
and C.~F.\ McKee, Phys. Fluids, {\bf 19}, 1130 (1976).

\bibitem[Gruzinov(2000)]{Gruzinov:00} A. Gruzinov, 
astro-ph/0012364 (2000).

\bibitem[Goodman and MacFadyen(2007)]{Goodman:07} J. Goodman, and 
A.~I.\ MacFadyen, arXiv:0706.1818 (2007);
L.\ Sironi, and J.\ Goodman, arXiv:0706.1819 (2007).


\bibitem[Meneguzzi et al.(1981)]{Meneguzzi:81} M. Meneguzzi,
U. Frisch, and A.\ Pouquet, Phys. Rev. Lett., {\bf 47}, 1060 (1981);  
A.~A.\ Schekochihin, {\it et al.}, Astrophys. J., {\bf 612}, 276 (2004).

\bibitem[Gruzinov(2007)]{Gruzinov:07} A. Gruzinov, arXiv:0704.3081 (2007).

\end{thebibliography}
\end{document}